\documentstyle[12pt,aaspp4,psfig]{article}

\def\grb{GRB\thinspace{990506}}
\def\vla{VLA\thinspace{J115450.1$-$2640.6}}
\def\ts{\thinspace}

\def\ale{\mathrel{\hbox{\rlap{\hbox{\lower4pt\hbox{$\sim$}}}\hbox{$<$}}}}
\def\age{\mathrel{\hbox{\rlap{\hbox{\lower4pt\hbox{$\sim$}}}\hbox{$>$}}}}

\begin{document}

\slugcomment{Accepted to the Astrophysical Journal Letters}         
\title{~~\\ ~~\\\large \bf The Rapidly Fading Afterglow from the \\ Gamma-Ray Burst of 1999 May 6\altaffilmark{1}}
 
\author{ 
G. B. Taylor\altaffilmark{2}, 
J. S. Bloom\altaffilmark{3}, 
D. A. Frail\altaffilmark{2}, 
S. R. Kulkarni\altaffilmark{3},\\
S. G. Djorgovski\altaffilmark{3}, \&
B. A. Jacoby\altaffilmark{3}
}

\altaffiltext{1}{The observations reported herein were obtained with the 
Very Large Array (VLA) operated by the National Radio Astronomy
Observatory which is a facility of the National Science Foundation
operated under a cooperative agreement by Associated Universities,
Inc.; and with the W. M. Keck
  Observatory, which is operated by the California Association
for Research in Astronomy, a scientific partnership among California
Institute of Technology, the University of California and the National
Aeronautics and Space Administration.}
\altaffiltext{2}{National Radio Astronomy Observatory, P.~O.~Box O,
  Socorro, NM 87801; gtaylor@nrao.edu, dfrail@nrao.edu}
\altaffiltext{3}{California Institute of Technology, 
Owens Valley Radio Observatory
105-24, Pasadena, CA 91125; jsb@astro.caltech.edu,
srk@astro.caltech.edu,
george@oracle.caltech.edu, baj@astro.caltech.edu}
 
\slugcomment{Accepted to the Astrophysical Journal Letters}  

\begin{abstract}
 
  We report on the discovery of the radio afterglow from the gamma-ray
  burst (GRB) of 1999 May 6 (GRB 990506) using the Very Large Array (VLA).
  The radio afterglow was detected at early times ($\Delta t=1.5$
  days), but began to fade rapidly sometime between 1 and 5 days after
  the burst.  If we attribute the radio emission to the forward shock
  from an expanding fireball, then this rapid onset of the decay in the
  radio predicts that the corresponding optical transient began to
  decay between 1 and 5 minutes after the burst. This could explain
  why no optical transient for GRB 990506 was detected in spite of
  numerous searches.  The cause of the unusually rapid onset of the
  decay for the afterglow is probably the result of an isotropically
  energetic fireball expanding into a low density circumburst
  environment.  At the location of the radio afterglow we find a faint
  ($R \sim 24$ mag) host galaxy with a double morphology.  
 
\end{abstract}

\keywords{gamma rays:bursts -- radio continuum:general --
  cosmology:observations}
 
\section{Introduction}

\grb\ was detected by the BATSE (trigger No.~7549) on board the
Compton Gamma-Ray Observatory (CGRO) on 1999 May 6.47 UT
(\cite{kip99a}). It was a bright burst, lasting approximately 150 s,
with a fluence in the 50-300 keV band of 2.23$\times10^{-4}$ erg
cm$^{-2}$ -- placing it in the top 2\% of BATSE bursts as ranked by
fluence. The PCA instrument on board the Rossi X-Ray Timing Explorer
(RXTE) began scanning a 8$^\circ\times6^\circ$ region centered on the
BATSE position some three hours after the burst (\cite{mt99}) and
discovered a previously uncataloged X-ray source, which faded
significantly over the observing interval (\cite{tm99}).  The initial
12\arcmin\ error circle for the RXTE-PCA position of \grb\ was further
refined to a 30 arcmin$^2$ region through the addition of timing
annuli from the Interplanetary Network (IPN), involving the NEAR,
Ulysses and CGRO satellites (\cite{hkkc99}).

Despite the fact that at least nine different optical telescopes
imaged the IPN/RXTE-PCA region containing the X-ray transient (some as
early as 90 minutes after the burst), no corresponding optical
transient was detected. In this paper we report the discovery of a
radio transient (\vla) in this field, which we propose is the radio
afterglow of GRB 990506.  We also present upper limits on the
brightness of the corresponding optical afterglow.  We discuss the
radio light curve of this transient, and its implications for the lack
of detection of the optical afterglow.  Finally, we present the
discovery of the probable host galaxy of the radio transient, and by
implication the GRB itself.  \grb\ joins a small but growing class of
bursts which evidently lack a bright optical transient but are seen at
both X-ray and radio wavelengths.

\section{Discovery of the Radio Afterglow}\label{discovery}

Very Large Array (VLA) observations were initiated 1.5 days after the
$\gamma$-ray burst. Details of this and all subsequent VLA
observations are given in Table~\ref{tab:Table-VLA}. In order to image
the entire 11.5\arcmin$\times$2.5\arcmin\ error region with the VLA at
8.46 GHz two pointings were required, each with a field-of-view at
half power response of 5.3\arcmin\ in diameter. In the southern field
we detect two sources whose flux density ($\sim{150}$ $\mu$Jy) did not
vary significantly from 1999 May 8 to 1999 May 21.  There were two
additional sources in the northern pointing, only one of which was
within the IPN/RXTE-PCA error box. This source, hereafter \vla, was
clearly detected during the first two epochs but had faded by at least
a factor of three thirteen days later. \vla\ was located at (epoch
J2000) $\alpha$\ =\ $11^h54^m50.13^s$ ($\pm{0.02^s}$) $\delta$\ =\ 
$-26^\circ40^\prime35.0^{\prime\prime}$ ($\pm{0.4^{\prime\prime}}$),
where the errors are the quoted 1$\sigma$ uncertainties in the
Gaussian fit obtained from the combined images on May 8 and May 9. The
source has not been detected since at any frequency (see Fig.~1).
A search was
  made for polarized emission by imaging the combined datasets on May
  8 and 9 in all four Stokes parameters. The 3$\sigma$ upper limits
  were 30\% for both linear and circular polarization.

Of immediate concern is the possibility that \vla\ is a background
variable radio source lying by chance within the moderately large
IPN/RXTE-PCA error box.  On the basis of the radio data alone we cannot
eliminate this possibility altogether.  However,
the order of magnitude change in source strength over the two weeks
immediately following the GRB, and the failure to detect any
subsequent radio emission after monitoring for 375 days, strongly
suggests that \vla\ was the afterglow of \grb.  We now proceed with
this hypothesis.

Reaching a flux density of 580 $\mu$Jy, \grb\ is in the top half of
the dozen radio radio afterglows detected to date (ranked by flux
density), fainter only than GRB\,970508, GRB\,980703, GRB\,991208, and
GRB\,991216. There is some indication that the flux density increased
from $\Delta t$=1.66 to 2.70 days (see Table~\ref{tab:Table-VLA}) but
the change in flux $\Delta$S=134$\pm$68 $\mu$Jy is only marginally
significant.  Some time between 2.5 days after the burst and 15.5
days, the radio afterglow declined below a detectable level. If we
describe this decline in flux density, $F_R$, as a power-law 
in time ({\it i.e.,} $F_R\propto
t^{\alpha_d}$) then the minimum slope of the decline depends on the
poorly determined time, $t_m$, for the onset of the decline such
that $\alpha_d^{-1} \simeq \log t_m - \log\ 15.66$. 



\section{Limits on the Optical Afterglow, and Discovery of the Host Galaxy}\label{ohost}

We obtained our first optical image of the field of GRB 990506 using
the 60-inch CCD Camera (160 arcmin$^2$ field of view) approximately 18
hr after the GRB and continued for the next four nights.  More details
of the R-band 
optical observations to date are presented in Table \ref{fig:log}.  No
confirmed optically variable sources in the IPN error box were found
by our group and others ({\it e.g.,}~Henden {\it et
  al.}~1999\nocite{hen+99}, \cite{phj+99a}).

At the position of the radio transient, an upper limit as strict as $R
\age 23.5$ (Pedersen {\it et al.}~1999b\nocite{phj+99b}) is placed on
the brightness of any point-source optical afterglow starting 12 hours
after the burst.  Table \ref{fig:log} summarizes the $R$-band optical
limits; our $V$-band observations on the first fours days also
revealed no counterpart to less stringent flux densities.

On 11 June 1999 UT, 35.78 days after the GRB, we re-observed the field
of GRB 990506 at the position of the radio transient using the Low
Resolution Imaging Spectrometer (LRIS; Oke {\it et
  al.}~1995\nocite{occ+95}) on the Keck II 10-meter Telescope on Mauna
Kea, Hawaii.  Six images of the field totaling 26 minutes of
integration were obtained in the Cousins $R$-band.  Observations of
the standard field PG 1323-086 (\cite{lan+92}) were used for magnitude
zero-point calibration, which we also checked against the photometry
by Vrba {\it et al.} (1999).  An astrometric plate solution was obtained
relative to the USNO A2.0 catalog (Monet {\it et al.}~1998) with a
statistical error of 0.24, 0.27 arcsec ($\alpha$, $\delta$).

Figure \ref{fig:opt} depicts the position of the radio transient and
persistent 1.4 GHz emission overlaid atop the optical Keck image.
Coincident with the position of the fading radio source is a faint
extended (NE-SW) galaxy with an irregular morphology.  The putative
host appears as two knots of roughly equal brightness, with the radio
transient position lying within the southwest knot.  We find $R = 24.0
\pm 0.3$ and $R = 24.4 \pm 0.3$ mag for the northeast and southwest
knots, respectively, based on photometry from the USNO (Vrba {\it et
  al.}~1999). The error in these magnitudes reflects the uncertainty
in the color of the galaxy and aperture correction.  The morphology of
the host is suggestive of a merging or interacting galaxy pair.

\section{Field radio Galaxies}

In a deep radio image taken at 1.4 GHz on 1999 June 12, two weak
background radio sources are seen within 30$^{\prime\prime}$ of \vla\ 
(Fig.~2).  The stronger of the two sources (R1) is 23$^{\prime\prime}$ to
the SE and has a flux density of 540 $\mu$Jy at 1.4 GHz.  This source
is undetected at 8.46 GHz, requiring either that it is resolved out at
higher resolution or that it has a moderately steep spectral index,
$\alpha < -1.1$, where $S \propto \nu^{\alpha}$.  The weaker of the
two nearby sources (R2) is detected some 30$^{\prime\prime}$ to the NW of
\vla.  This source has a flux density of 170 $\mu$Jy at 1.4 GHz and a
spectral index steeper than $-0.5$. From recent spectroscopic
observations (see table \ref{fig:log}) we identify the two radio
sources (R1 and R2) with an early-type spiral and a QSO at redshifts $z =
0.326$ and $z = 0.273$, respectively.

\section{Discussion}

The simplest explanation for the behavior of the radio afterglow from
\grb\ is that the emission originates in the forward shock driven into
the surrounding medium by the relativistically expanding blast wave
(\cite{spn98}). In this case the apparent rise in the 8.46 GHz light
curve to a maximum $F_m$, followed by a decay is the result of an
evolving synchrotron spectrum, produced from electrons accelerated in
the shock, whose peak frequency $\nu_m$ passed through the band during
our observing interval. As noted in \S\ref{discovery}, the exact
time, $t_m$, for the onset of the decay and the value of the
power-law slope, $\alpha_d$, are uncertain owing to the undersampling
of the light curve. Given the data, a range of $t_m$ values from 1 to
5 days are possible. The upper limit on $t_m$ is determined by
requiring that $\alpha_d>-2$.  Significantly steeper values have yet
to be seen for other afterglows.  Furthermore, this value is entirely
consistent with the X-ray decay $\alpha_d=-1.9 \pm 0.6$ for this burst
as measured by the RXTE-ASM (\cite{tm99}). For $1\leq t_m \leq 5$ days
the corresponding values of the decay slope are $-0.8 \leq \alpha_d
\leq -2$.  

\begin{figure*} 
\centerline{\hbox{\psfig{figure=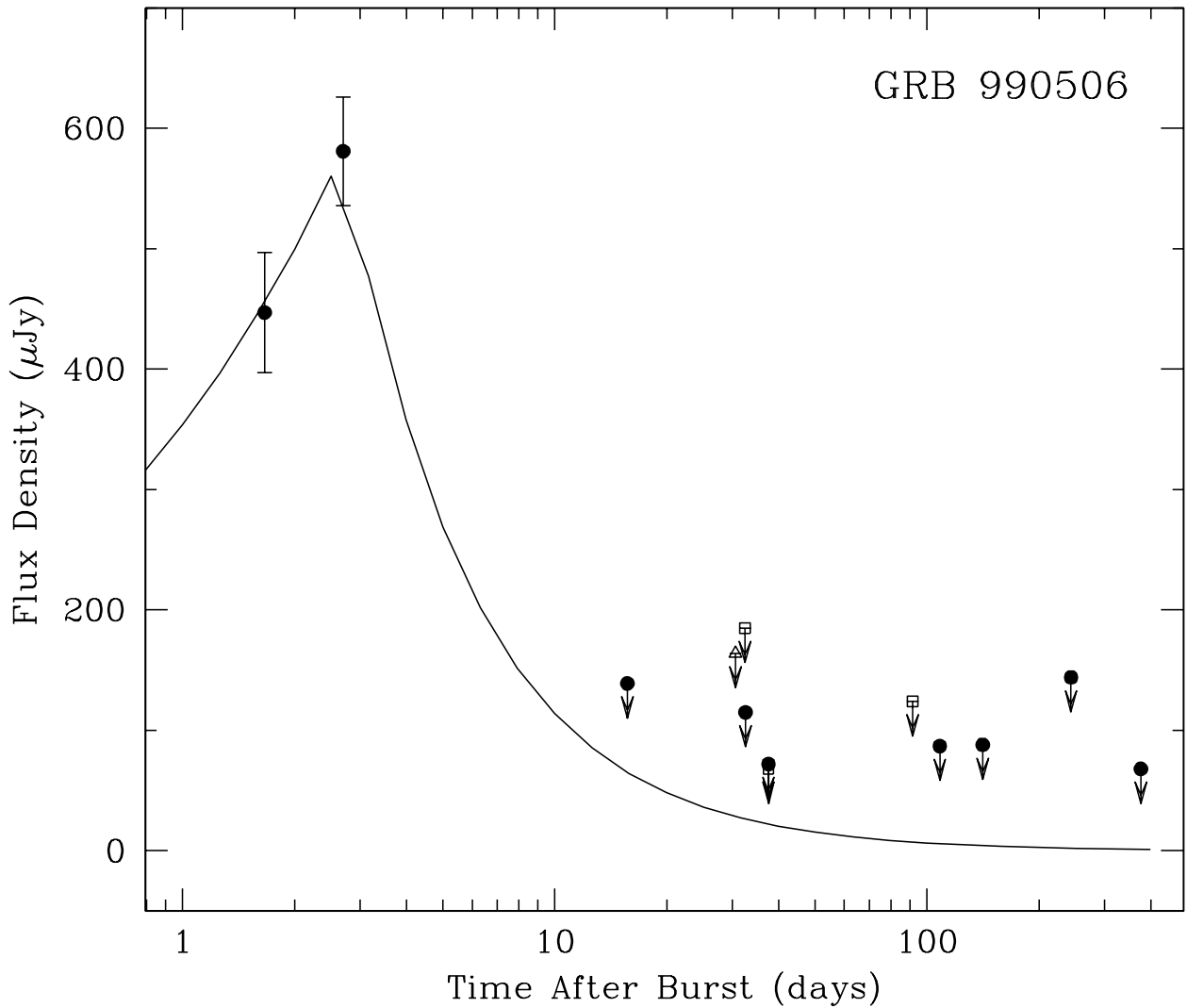,width=7.0in}}} 
\figcaption{The radio light curve for \vla.  Key to symbols: filled circles
  are 8.46 GHz observations; open squares are 1.4 GHz observations and
  the open triangle represents a single observation at 4.86 GHz. Upper
  limits are ploted as the flux density at the position of \vla\ (if
  positive) plus
  2.05 times the rms noise (see Table~\ref{tab:Table-VLA}). The solid
  line is the expected emission at 8.46 GHz for a forward shock
  propagating into a constant density medium. The light curve rises as
  $t^{1/2}$ until $t_m$=2.5 days, and thereafter decays with
  $\alpha_d=-1.25$ ({\it i.e.,} $t^{\alpha_d}$).  The model is meant to be
  representative only.  There is a range $t_m$ and $\alpha_d$,
  allowed by these data (see text for details).
\label{fig:rlight}}
\end{figure*}

\begin{figure}
\centerline{\hbox{\psfig{figure=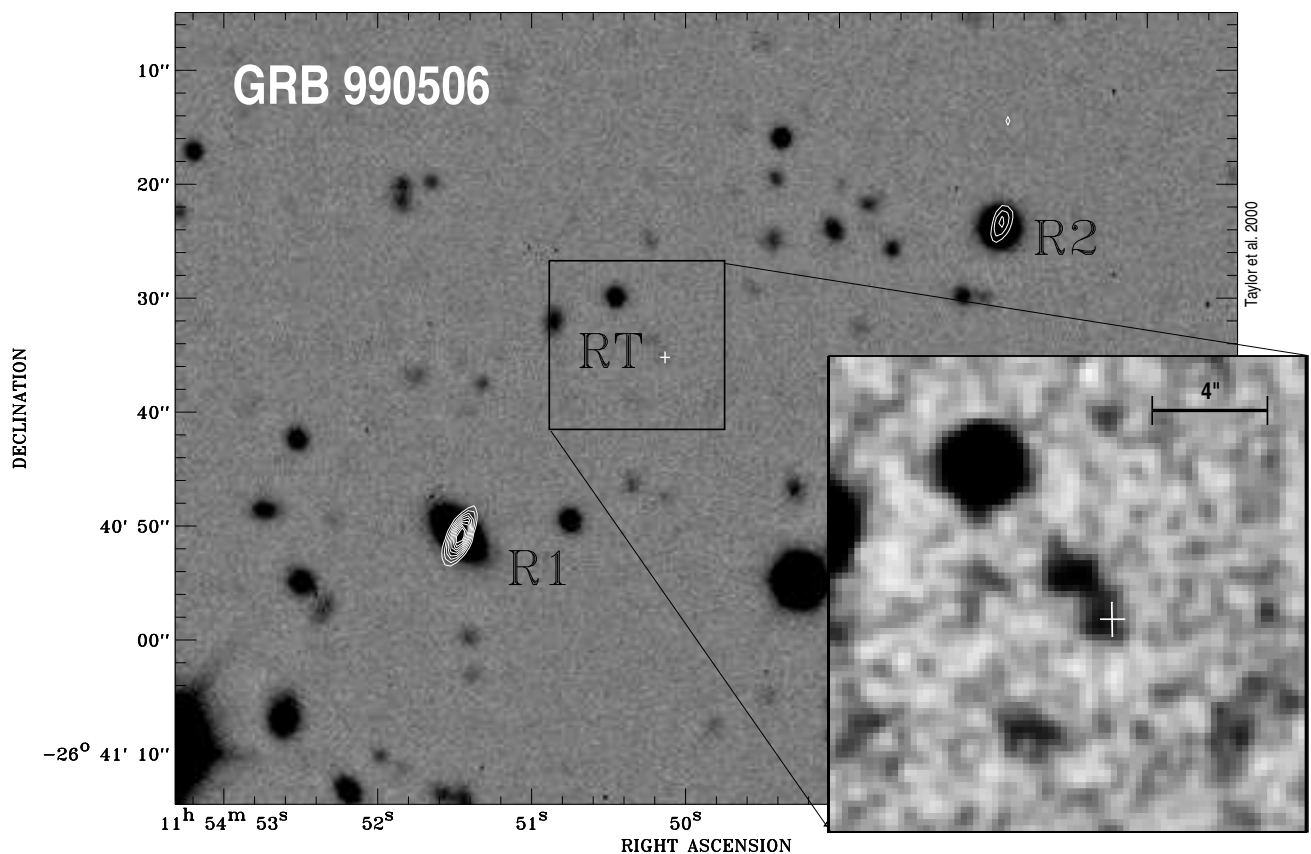,width=7.0in,angle=-90}}} 
\figcaption{The optical field of GRB 990506 with radio overlay.  The
optical image a sum of 1500 s from Keck taken 35 days after the GRB.
The radio contours (white) are in incremental units of r.m.s.~noise
(2--10 $\sigma$) at an observing frequency of 1.4 GHz.  The cross shows the relative
2-$\sigma$ position of the transient radio source associated with GRB
990506.  Added in quadrature to the radio transient positional uncertainty are
the r.m.s.~statistical uncertainty of the optical tie to the USNO A2.0
catalogue (see text) and the systematic error in the tie of USNO A2.0
and the International Coordinate Reference frame (Deutsch
1999).  The radio transient is coincident with the
south-western knot of a faint galaxy, whose bimodal morphology is 
suggestive of an interacting system.\label{fig:opt}
}
\end{figure}

In Figure \ref{fig:rlight} we plot a 8.46 GHz light curve
expected for
a adiabatic forward shock, propagating in a homogeneous medium to
demonstrate that such models are consistent with the data.  In this
representative model, the peak flux is 580 $\mu$Jy, $t_m$=2.5 days,
and $\alpha_d=-1.25$. The tightest constraint on $\alpha_d$ and/or
$t_m$ comes from the 1.4 GHz flux density upper limit on 1999 June
12.92 UT.  Reasonable model inputs predict peak flux densities at this
frequency of order 100-200 $\mu$Jy. Therefore, at least this one epoch
favors the steeper decay indices and the smaller $t_m$ values from
the ranges given above.

The estimated $t_m$ value suggests that \grb\ entered the decay phase
at 8.46 GHz considerably earlier than previous radio afterglows.  At
this same frequency the decay timescales for GRB\,970508
(\cite{fwk00}), GRB\,980329 (\cite{fra+00a}), GRB\,980519
(\cite{fra+00b}), GRB\,990510 (\cite{hbf+99}), 
GRB\,981216 (\cite{fkb+99}) range from 8.5 days to 90 days.  Only
GRB\, 990123 exhibited a faster decay (\cite{kfs+99}; see below).

One consequence of the early radio decay is that the apparent absence
of the optical afterglow can now be understood rather simply without
the need to invoke excess extinction along the line of sight. If the
synchrotron peak moved through the radio band $\nu_R$ on the order of
$t_m$=1-5 days, then the corresponding timescale for the peak to move
through the optical bands $\nu_o$ is $t_o=t_m\,(\nu_R/\nu_o)^{2/3}$,
or $60\leq t_m \leq 300$ s (\cite{mr97a}). This small timescale can be
contrasted with GRB\thinspace{971214} for which Ramaprakash {\it et
  al.} (1998)\nocite{rkf+98} derived $t_o$=0.6 days at
$\nu_m=3\times{10}^{14}$ Hz. In general, the flux density decays from
its peak $F_m$ as a power law with index $\alpha_d$ for $t>t_o$.
Provided that the synchrotron peak frequency $F_m$ remains constant
(as it should if dealing with a spherical, adiabatic expansion), then
the peak 8.4 GHz flux density of 580 $\mu$Jy predicts a peak optical
afterglow emission of $R = 17$ mag (after correcting for foreground
extinction).  Even a relatively shallow decay, $\alpha_d = -1.25$,
predicts that the OT was already undetectable at $R=22$ mag 90 minutes
after the burst when the field of GRB 990506 was observed by Zhu \&
Zhang (1999).  By the time of the deep $R$-band observations ({\it
  e.g.,} \cite{mpp+99}, \cite{phj+99b}) the OT had $R \age 25$ mag and was
likewise undetectable.

Perhaps more importantly, the early radio decay is indicative of
unusual physical conditions. Short-lived but moderately bright
afterglows can result if the GRB exploded with a high energy
($E_\circ$) into a low density medium ($n_\circ$), or, as noted by
Galama {\it et al.} (1999)\nocite{gbw+99}, the magnetic field in the forward
shock was weak. Unfortunately, while we have constrained $F_m$ and
$\nu_m(t_m)$, the paucity of broadband data for this burst does not allow
us to determine other important observables such as the synchrotron
self-absorption frequency $\nu_a$ and the cooling frequency $\nu_c$.
Without this information it is not possible to fully constrain the
physical properties of the afterglow and the surrounding medium
(\cite{wg99}).


Alternatively, prompt radio emission, analogous to the well-known
optical emission (\cite{abb+99}), was also seen in the afterglow from
GRB\ts{990123} (\cite{kfs+99}) and perhaps GRB\,970828 as well. This
radio emission arises in the reverse shock in the days following a
$\gamma$-ray burst (\cite{sp99b}). It is suppressed at early times by
synchrotron self-absorption, while at later times the light curve
decays rapidly ($\propto{t^{-2}}$).


We note that it is possible that the southwest knot may itself be the
optical transient since the GRB field has not been reimaged since 35
days after the burst.  However, even a shallow decay in the optical of
$t^{-1}$ then predicts an $R$-band magnitude of 20.1 one day after the burst
which was not observed.  We thus find it unlikely that the southwest
knot could be an optical transient given the early non-detections.
Instead, we interpret the two knots as part of the host galaxy of GRB
990506.  The apparent $R$-band magnitude of the putative host of GRB
990506 is similar to that of other GRB host galaxies.  Spectroscopic
observations are necessary in order to establish the redshift and
other physical properties of this host galaxy.  The bimodal morphology may be
indicative of merger activity.  In this respect the host
appears similar to a growing number of GRB hosts with irregular
morphology: GRB 980613 (Djorgovski {\it et al.}~1999\nocite{djo+99}),
GRB 970828 (Djorgovski {\it et al.}~2000\nocite{djo+00}), and GRB
990123 (Bloom {\it et al.}~1999\nocite{bod+99}).

\section{Conclusions}

We have identified the radio afterglow from GRB 990506.  This GRB was
unusual in that it produced a radio afterglow that began to fade at
very early times (between 1 and 5 days after the burst) and in that no
optical afterglow was detected in spite of the numerous deep images
obtained.  Both these observations may be explained by a high energy
spherical fireball expanding into a low density environment.  In this
simple picture there is no need to invoke dust extinction to account
for the lack of detection of an optical afterglow.  We cannot rule
out, however, that the radio emission originated in a reverse shock.
If the reverse shock produced the radio afterglow, then the emission
from the forward shock (in both optical and radio) was presumably too
faint to be seen.  To find additional GRBs of this type will require
more rapid followup to precise burst localizations than has typically
been achievable.  Upcoming satellite missions such as HETE II and
SWIFT should improve upon this situation.

\acknowledgements

We are grateful to D. Frayer, A.  Eichelberger, G. Oelmer for
observations at Palomar. W.~W.~Sargent, T.  Small, A. Diercks, and
T.~J.~Galama are thanked for their contribution at Keck.  JSB
gratefully acknowledges support from the Fannie and John Hertz
Foundation.  SRK's research is supported by grants from NSF and NASA.
SGD acknowledges partial support from the Bressler Foundation.

\clearpage
  
 
\bibliographystyle{../g991216/apj1b}

\clearpage
 
\newpage 
\begin{deluxetable}{lrcrr}
\tabcolsep0in\footnotesize
\tablewidth{\hsize}
\tablecaption{VLA Observations of \grb\ \label{tab:Table-VLA}}
\tablehead {
\colhead {Epoch}      &
\colhead {$\Delta t$} &
\colhead {Freq.} &
\colhead {$F_R\pm\sigma$} &
\colhead {98\% confidence} \\
\colhead {(UT)}      &
\colhead {(days)} &
\colhead {(GHz)} &
\colhead {($\mu$Jy)} &
\colhead {($\mu$Jy)}
}
\startdata
1999 May  $\phantom{0}$8.13 & 1.66   & 8.46  & 447$\pm$50 &    \nl
1999 May  $\phantom{0}$9.17 & 2.70   & 8.46  & 581$\pm$45  &   \nl
1999 May  22.13             & 15.66  & 8.46  &  55$\pm$41  &  $<$139   \nl
1999 Jun. $\phantom{0}$6.09 & 30.62  & 4.86  & $-$46$\pm$80  &  $<$164 \nl
1999 Jun. $\phantom{0}$7.96 & 32.49  & 1.40  & $-$21$\pm$90 &  $<$185  \nl
1999 Jun. $\phantom{0}$8.03 & 32.56  & 8.46  &  41$\pm$36   &  $<$115   \nl
1999 Jun. 12.99             & 37.52  & 8.46  &  8$\pm$31  &  $<$72 \nl
1999 Jun. 13.00             & 37.53  & 1.40  & $-$80$\pm$33 &  $<$68  \nl
1999 Aug. $\phantom{0}$6.03 & 91.56  & 1.40  & 1$\pm$60   &  $<$124 \nl
1999 Aug. 22.82             & 108.35  & 8.46  & 25$\pm$30  &  $<$87 \nl
1999 Sep. 24.70             & 141.23  & 8.46  & 37$\pm$25  &  $<$88 \nl
2000 Jan. $\phantom{0}$5.50             & 244.03  & 8.46  & $-$54$\pm$70 &  $<$144  \nl
2000 May. 16.16             & 375.69  & 8.46  & $-$5$\pm$33 &  $<$68  \nl
\enddata
\tablecomments{The columns are (left to right), (1) UT date of the
  start of each observation; (2) Time elapsed since the $\gamma$-ray
  burst; (3) The observing frequency; (4) The peak flux density at
  the best fit position of the radio transient, with the error given
  as the root mean square (rms) flux density in the off-source region of the 
  image; and (5) The 98\% confidence upper limit computed from 
the measured value in (4) if greater than 0 plus 2.05 times the rms.  
Observations employed a
  bandwidth of 100 MHz and all four Stokes parameters were recorded.
  The instrumental phase was calibrated with J1146$-$247 and
  J1145$-$228 and the flux scale was tied to J1331+305.}
\end{deluxetable}

\begin{table*}[tb]
\caption{Optical Imaging and Spectroscopic Observations of GRB 990506}
\label{fig:log}
\vskip 12pt
\begin{tabular}{lcccccccr}
\hline\hline
\multicolumn{1}{c}{Date}& Inst.$^a$ & 
\multicolumn{1}{c}{Int.~Time} & $\Delta t$ & 
\multicolumn{1}{c}{$R$ Mag$^b$} & Reference\\

\multicolumn{1}{c}{UT} &  & (sec) & (days) &
\multicolumn{1}{c}{observed} \\ 
\hline
1999 May 6.53 & BAO & 300 & 0.058 & $\age 19.0$ & Zhu \& Zhang
1999\nocite{zhu+99} \\
1999 May 6.96 &NOT  & 480 & 0.486  & $\age 23.5$ & Pedersen {\it et
  al.} 1999b\nocite{phj+99b}\\
1999 May 7.228&P60  & 1920 & 0.754 & $\age 22.1$ & this paper \\
1999 May 7.565 & ESO & 1800 & 1.019 &$\age 23.0^{c}$ & 
Masetti {\it et al.}~1999\\  
1999 May 8.231&P60  & 1920 & 1.757 & $\age 22.5$ & this paper \\
1999 May 8.284&P200 & 600  & 1.81  & $\age 22.6$ & this paper \\
1999 May 9.226&P60  & 1920 & 2.752 & $\age 20.9$ & this paper \\
1999 Jun 11.25&LRIS & 1500 & 35.78 & 24.4 $\pm$ 0.3 & this paper\\
\hline

2000 Feb 14.49 & ESI & 5400  & 285.02 & \ldots & this paper \\

\hline
\end{tabular}
\raggedright

{\footnotesize 

\noindent $^a$ Instruments: BAO = 0.6/0.9m Schmidt telescope in 
Xinglong; NOT = The ALFOSC on the 2.56-m Nordic
Optical Telescope; P60 = CCD Camera on the 60-inch Palomar Telescope;
P200 = COSMIC Reimaging on the 200-inch Hale Telescope, Palomar; LRIS
= Low-Resolution Spectrometer on 10-m Keck II, Mauna Kea. ESI =
Echellete Spesctrograph and Imager on the Keck II 10-m.

\noindent $^b$ Upper limits are 3$\sigma$ for a point-source
detection at the position of the radio afterglow unless otherwise noted. 

\noindent $^{c}$ This value is a 3$\sigma$ upper limit to variability in the IPN
   localization greater than $R = 0.3$ mag as compared to later
   epochs. 
}
\end{table*}

\clearpage 

\clearpage 

\end{document}